\documentclass[aps,prl,reprint]{revtex4-2}

\usepackage{amsmath}
\usepackage{amssymb}
\usepackage{bm}
\usepackage{xcolor}
\usepackage{subcaption}
\usepackage{placeins}
\usepackage{float}
\usepackage[english=british]{csquotes}

\usepackage{graphicx}
\usepackage[a4paper,left=1.5cm,right=1.3cm,top=3.0cm,bottom=2cm]{geometry}

\renewcommand{\Pr}   {\ifmmode \mathrm{Pr}  \else $\mathrm{Pr}$ \fi} 
\newcommand{\Ra}     {\ifmmode \mathrm{Ra}  \else $\mathrm{Ra}$ \fi} 
\newcommand{\Bi}     {\ifmmode \mathrm{Bi}  \else $\mathrm{Bi}$ \fi} 
\newcommand{\RaN}    {\ifmmode \mathrm{Ra}_{\textrm{N}}  \else $\mathrm{Ra}_{\textrm{N}}$ \fi} 
\newcommand{\RaD}    {\ifmmode \mathrm{Ra}_{\textrm{D}}  \else $\mathrm{Ra}_{\textrm{D}}$ \fi} 
\newcommand{\Racrit} {\ifmmode \mathrm{Ra}_{\textrm{crit}}  \else $\mathrm{Ra}_{\textrm{crit}}$ \fi} 
\newcommand{\Ro}     {\ifmmode \mathrm{Ro}  \else $\mathrm{Ro}$ \fi} 
\newcommand{\Ta}     {\ifmmode \mathrm{Ta}  \else $\mathrm{Ta}$ \fi} 
\newcommand{\Ek}     {\ifmmode \mathrm{Ek}  \else $\mathrm{Ek}$ \fi} 
\newcommand{\Nu}   		{\mathrm{Nu}}
\newcommand{\NuN}    {\ifmmode \mathrm{Nu}_{\textrm{N}}  \else $\mathrm{Nu}_{\textrm{N}}$ \fi} 
\newcommand{\NuD}    {\ifmmode \mathrm{Nu}_{\textrm{D}}  \else $\mathrm{Nu}_{\textrm{D}}$ \fi} 
\newcommand{\NulocT}    {\ifmmode \mathrm{Nu}_{\textrm{loc}, T}  \else $\mathrm{Nu}_{\textrm{loc}, T}$ \fi} 
\newcommand{\Nuloc}    {\ifmmode \mathrm{Nu}_{\textrm{loc}}  \else $\mathrm{Nu}_{\textrm{loc}}$ \fi} 
\newcommand{\Nulocexp}    {\ifmmode \mathrm{Nu}_{\textrm{loc}}^{\textrm{exp}}  \else $\mathrm{Nu}_{\textrm{loc}}^{\textrm{exp}}$ \fi} 

\newcommand{\TN}    {\ifmmode T_{\textrm{N}}  \else $T_{\textrm{N}}$ \fi} 
\renewcommand{\Re}   {\ifmmode \mathrm{Re}  \else $\mathrm{Re}$ \fi} 
\newcommand{\dT}     {\ifmmode \Delta T  \else $\Delta T$ \fi} 
\newcommand{\dTD}    {\ifmmode \Delta T_{\textrm{D}}  \else $\Delta T_{\textrm{D}}$ \fi} 
\newcommand{\dTN}    {\ifmmode \Delta T_{\textrm{N}}  \else $\Delta T_{\textrm{N}}$ \fi} 
\newcommand{\Fourier}{\ifmmode \mathcal{F}  \else $\mathcal{F}$ \fi} 
\newcommand{\real}   {\ifmmode \mathfrak{R} \else $\mathfrak{R}$ \fi} 
\newcommand{\imag}   {\ifmmode \mathfrak{I} \else $\mathfrak{I}$ \fi} 

\begin{document}




\title{Thermal boundary condition studies in large aspect ratio Rayleigh-Bénard convection}
\author{Theo K\"aufer$^1$, Philipp P. Vieweg$^1$, J\"org Schumacher$^{1,2}$, and Christian Cierpka$^1$}

\affiliation{$^1$ Institute of Thermodynamics and Fluid Mechanics, Technische Universit\"at Ilmenau, D-98684 Ilmenau, Germany,\\
                $^2$ Tandon School of Engineering, New York University, New York City, NY 11201, USA}
%




\begin{abstract}
We study the influence of thermal boundary conditions on large aspect ratio Rayleigh-Bénard convection by a joint analysis of experimental and numerical data sets for a Prandl number $\Pr =  7$ and Rayleigh numbers $\Ra = 10^5 - 10^6$. The spatio-temporal experimental data are obtained by combined Particle Image Velocimetry and Particle Image Thermometry measurements in a cuboid cell filled with water at an aspect ratio $\Gamma = 25$. In addition, numerical data are generated by Direct Numerical Simulations (DNS) in domains with $\Gamma = 25$ as well as $\Gamma = 60$ subject to different thermal boundary conditions. Our experimental data show an increased \textit{characteristic horizontal extension scale} of the flow structures, $\tilde{\lambda}$, for increasing $\Ra$, which is coupled with a raise of the Biot number $\Bi$ in particular at the cooling plate. However, we find the experimental flow structure size to range in any case between the ones observed for the idealized thermal conditions captured by the simulations. On the one hand, they are larger than in the numerical case with applied uniform temperatures at the plates, but, on the other hand, smaller than in the case of an applied constant heat flux, the latter of which leads to a structure that grows gradually up to the horizontal domain size. We are able to link this observation qualitatively to theoretical predictions for the onset of convection. Furthermore, we study the effect of the asymmetric boundary conditions on the \textit{heat transfer}. Contrasting experimental and numerical data reveals an increased probability of far-tail events of reversed heat transfer. The successive decomposition of the local Nusselt number $\Nuloc$ traces this effect back to the sign of the temperature deviation $\tilde{\Theta}$, eventually revealing asymmetries of the heating and cooling plate on the thermal variance of the generated thermal plumes. 
\end{abstract}

\clearpage

\maketitle

\section{Introduction}
\label{Sec:Introduction}
Thermal convection drives many natural flow phenomena such as convection in the Earth's outer core \cite{Guervilly2019, Finlay2011}, mantle \cite{Christensen1995, Zhong2000}, ocean \cite{Maxworthy1994, marshall1999open}, or atmosphere \cite{Atkinson1996, mapes1993cloud}. Beside its presence throughout the layers of Earth, convection can also be found, e.g., in the Sun and other planets of our solar system \cite{Schumacher.2020Colloquium, Hanson2020, Young.2017Forward, GarciaMelendo.2013Atmospheric}. Characteristic to all of these systems is a typically large aspect ratio $\Gamma=W/H$ \cite{Ahlers.2009Heat,Chilla.2012New}, i.e., the height of the flow domain $H$ in the direction of gravity is typically much smaller compared to the width $W$ perpendicular to it.

Albeit thermal convection has meanwhile been studied for more than 100 years \citep{Benard1901, Rayleigh1916}, the exploration of convection in large aspect ratios intensified just recently \cite{Pandey.2018Turbulent, StevensRichardJ.A.M..2018Turbulent, Krug.2020Coherence, Vieweg.2021Supergranule}. 
Thereby, the formation of long-living large-scale flow structures  -- in this case, the so-called turbulent superstructures, see also Section \ref{Sec:turb_superstructures} -- was discovered by numerical investigations of large aspect ratio Rayleigh-Bénard convection (RBC) \cite{Pandey.2018Turbulent,StevensRichardJ.A.M..2018Turbulent}. Since their discovery, these flow structures are subject of intense research due to their clear distinction from small-scale turbulence, which acts on much smaller time scales. Besides the effect of the thermal boundary condition, which we discuss in a general sense in Section \ref{Sec:turb_superstructures} and specifically for the case of mixed and asymmetric boundary conditions in the remainder of this paper, further aspects of the turbulent superstructures have been investigated. For example, the coherence of velocity and temperature structures was shown by analysis of the temperature and velocity co-spectra. \cite{Krug.2020Coherence}. Other studies focused on the interaction between small-scale fluctuations and turbulent superstructures \cite{Berghout.2021large,Green.2020Resolved}, underlining the richness of large aspect ratio convection.

As a complement to these studies in the Eulerian frame of reference, the role of large-scale flow structures on the Lagrangian or material transport has been revealed by means of spectral clustering of Lagrangian trajectories \cite{Schneide.2018Probing, Vieweg.2021Lagrangian, Schneide.2022Evolutionary}. Theoretical and numerical studies are supported and extended by experimental analyses of the long-term behavior of turbulent superstructures together with their local convective heat transfer based on spatially and temporally resolved temperature and velocity measurements \cite{Moller.2021Long, Moller.2022Combined}.

Contrasting experimental and numerical results, Moller et al. \cite{Moller.2022Combined} observed two striking disagreements: (i) the overall heat transport across the fluid layer is decreased, and (ii) the flow structures are significantly larger in the experimental setup compared to the numerical results. Interestingly, these discrepancies increase once the Rayleigh number $\Ra$ -- as a measure of the strength of the thermal driving, see eq. \eqref{eq:def_Pr_RaD_RaN} -- is increased. As the Prandtl number $\Pr$ -- defining the working fluid, see again eq. \eqref{eq:def_Pr_RaD_RaN} -- is constant and these two control parameters specify the entire dynamical system, this observation suggests that these observations are a feature of the experiment, which are not captured by idealized simulations.
Considering the experimental setup, Moller et al. \cite{Moller.2022Combined} attributed these disagreements to the non-ideal thermal boundary condition at the cooling plate made from glass. In the present work, we pick up these observations and analyze the influence of the present asymmetric thermal boundary conditions in more detail.

The impact of thermal boundary conditions on the heat transport across the fluid layer -- as can be quantified by the Nusselt number $\Nu$, see eqs. \eqref{eq:def_NuD}, \eqref{eq:def_NuN}, and \eqref{eq:Nu_loc_T} -- was investigated in several past studies \cite{Verzicco.2004Effects, Brown.2005Heat, VERZICCO.2008comparison, Johnston.2009Comparison, Foroozani.2021Turbulent, Vieweg.2021Supergranule} and led, on the one hand, to suggestions of corrections for the influence of the plates' materials \cite{Verzicco.2004Effects, Brown.2005Heat}, whereas, on the other hand, deviations from isothermal conditions seemed to accelerate the convective heat transport at least up to $\Ra \sim 10^{7}$ \cite{Johnston.2009Comparison, Foroozani.2021Turbulent, Vieweg.2021Supergranule}.
However, except for \cite{Vieweg.2021Supergranule}, none of these studies provided a horizontally sufficiently extended domain such that turbulent superstructures or equivalent structures could form \cite{Manneville2006, Cross2009, Koschmieder1993}. This circumstance becomes crucial, given the fact that the large-scale flow structures transport most of the heat across the fluid layer \cite{Krug.2020Coherence}.

These reasons underline the relevance of further investigations of the influence of mixed or asymmetric thermal boundary conditions on the heat transfer and large-scale flow structure formation in large aspect ratio domains. First, we contextualize the experimentally present conditions, investigate the size and evolution of the large-scale flow structures by extracting the characteristic wavelength from the power spectrum, and compare our observation with theoretical predictions for the onset of convection. Second, we analyze the local heat transfer and compare the corresponding experimental and numerical probability density functions (PDFs). We observe higher probabilities of a reversed heat transfer for experimental data. The successive decomposition of the local Nusselt number $\Nuloc$ eventually unveils the underlying effect of the asymmetric thermal boundary conditions.

The remainder of this manuscript is organized as follows. We start by outlining large-scale flow structures and their formation in Section \ref{Sec:turb_superstructures} and, thereupon, briefly describe the experimental and numerical methods in Section \ref{Sec:Methods}. We then address the large-scale flow structure size in Section \ref{Sec:superstructures}, whereas the heat transfer is analyzed in Section \ref{Sec:heat_transfer}. We conclude with a discussion of our results in Section \ref{Sec:Discussion}.

\section{Large-scale flow structures in thermal convection}
\label{Sec:turb_superstructures}
\begin{figure}[tb]
    \centering

    \includegraphics[width=\linewidth]{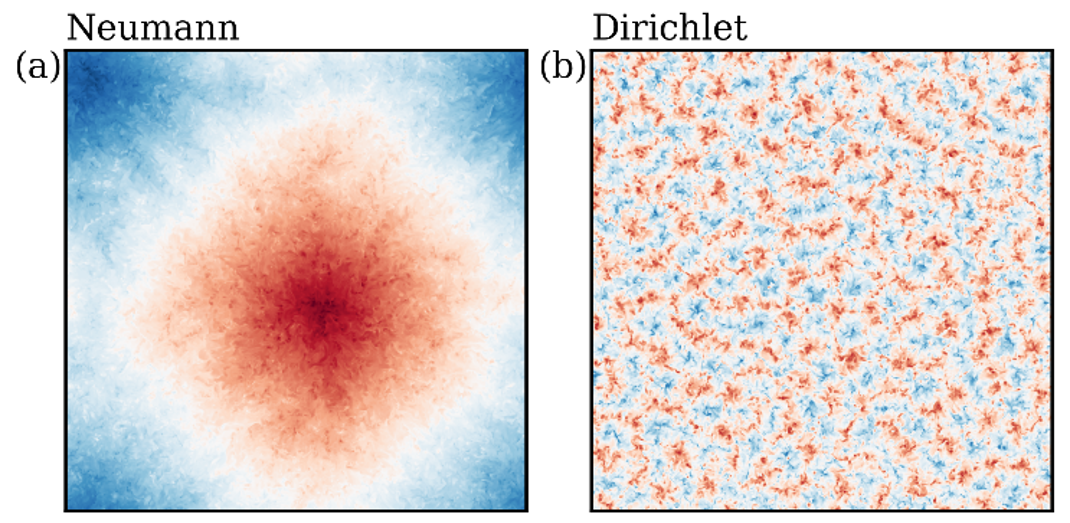}
    \centering
    \caption{Thermal boundary conditions determine the large-scale flow structure formation, the latter of which are termed \textit{supergranules} or \textit{turbulent superstructures} in the case of Neumann or Dirichlet conditions, respectively. The panels visualize the temperature fields in horizontal cross-sections at aspect ratio $\Gamma = 60$ from \cite{Vieweg.2021Supergranule}.}
    \label{fig:comparison_structures}   
\end{figure}
As briefly introduced above, Rayleigh-Bénard convection exhibits characteristic long-living large-scale flow structures (LLFSs) that set themselves apart from smaller-scale turbulence acting on significantly shorter time scales. While the latter capture smaller eddies that might be as small as the Kolmogorov scale $\eta_{\textrm{K}} \ll H$ and correspond to rapid temporal fluctuations, the former offer characteristic horizontal extensions with typical length scales of at least $\mathcal{O} \left( H \right)$ and time scales $\tau_{\textrm{LLFSs}} \gg \tau_{\textrm{f}}$ (with $\tau_{\textrm{f}}$ representing the convective time unit, see also Section \ref{Sec:Numerical_methods}). Due to this clear scale separation, these different processes or flow structures can be separated by a Reynolds decomposition, or temporal filter \cite{Pandey.2018Turbulent, Vieweg.2021Supergranule}.

Considering the simplest configuration of Rayleigh-Bénard convection with no additional physics such as rotation \cite{Takehiro2002, Stevens2013,Vieweg.2022Inverse}, magnetohydrodynamic effects \cite{Burr2001, Burr2002, Zurner2020}, or complex fluid property dependencies \cite{Valori2017, Yik2020, Pandey2021}, the nature of these long-living large-scale flow structures have recently been found to depend decisively on the aspect ratio \cite{Ahlers.2009Heat, Zurner2020, Pandey.2018Turbulent, StevensRichardJ.A.M..2018Turbulent, Foroozani.2021Turbulent} and thermal boundary conditions \cite{Foroozani.2021Turbulent, Vieweg.2021Supergranule}. If the aspect ratio is small with $\Gamma \approx 1$, a so-called \textit{large-scale circulation} or mean wind with a characteristic horizontal extension of about $H$ forms \cite{Ahlers.2009Heat} -- in this case, the large-scale flow structure is (independently of the thermal boundary conditions \cite{Foroozani.2021Turbulent}) clearly dictated by the lateral boundaries. This changes once the flow is released from the impact of the latter. Their importance decreases with $\mathcal{O} \left( \Gamma^{-2} \right)$ \cite{Cross2009, Koschmieder1993, Manneville2006}, so domains with $\Gamma \gtrsim 20 \gg 1$ can be considered being \enquote{large} and fairly close approximations of infinitely extended fluid layers \cite{Koschmieder1993, StevensRichardJ.A.M..2018Turbulent, Krug.2020Coherence}. As the impact of the lateral boundaries vanishes, so arise other, or new dependencies that govern the nature of the establishing long-living large-scale flow structures -- albeit discovered just recently \cite{Vieweg.2021Supergranule}, the thermal boundary conditions offer the perhaps most striking effect. 

In the classical Dirichlet case of applied uniform \textit{temperatures} at the plates, so-called \textit{turbulent superstructures} form with time-independent characteristic horizontal extensions of order $\mathcal{O} \left( H \right)$ \cite{Pandey.2018Turbulent, Vieweg.2021Supergranule} -- these structures are in clear contrast to the time-dependent \textit{gradual supergranule aggregation} that takes place in the opposite Neumann case of an applied uniform \textit{temperature gradient} (or heat flux) at the plates \cite{Vieweg.2021Supergranule}. More details on the mathematical aspects of these choices are provided in Section \ref{Sec:Numerical_methods}, whereas Section \ref{Sec:superstructures} will highlight that the particular choice is physically related to the propagation or relaxation of thermal perturbations in the plates and the fluid. Interestingly, this latter process leads (if not controlled by rotation \cite{Vieweg.2022Inverse}) to a growth of the structures until the horizontal domain size is reached, i.e., their characteristic length $\tilde{\lambda} \sim \Gamma \gg \mathcal{O} \left( H \right)$, exceeding thus the turbulent superstructures clearly in terms of their length scale. Figure \ref{fig:comparison_structures} (a, b) compares these turbulent long-living large-scale flow structures in a $\Gamma = 60$ domain for an equivalent heat transfer through the fluid layer. Crucially, other aspects such as the Rayleigh number, Prandtl number, or mechanical boundary conditions \cite{Pandey.2018Turbulent, Vieweg.2021Supergranule} become secondary once compared to this dominant effect of thermal boundary conditions. As these observations in the turbulent regime are qualitatively in accordance with the behavior at and slightly above the onset of convection, this highlights the importance of primary and secondary instabilities for the dynamical system in general. We will thus address the transition from Dirichlet to Neumann conditions in more detail in Section \ref{Sec:superstructures}.

Additional physical mechanisms, such as rotation around the vertical axis, may additionally influence the long-living large-scale flow structures. Vieweg et al.~\cite{Vieweg.2022Inverse} showed that the impact of rotation depends quite differently on the choice of the thermal boundary condition or nature of the latter. However, such additional effects are beyond the scope of the present manuscript with its focus on the thermal boundary conditions.

Experimental studies are the method of choice for long-time studies of turbulent superstructures since the expenses for direct numerical simulations in horizontally extended domains scale with $\sim \Gamma^{2}$ \cite{BailonCuba.2010Aspect} and $\sim t$, restricting numerical studies of the long-term behaviour of these long-living large-scale flow structures seriously. Given the slow reorganisation of the turbulent superstructures over time \cite{Pandey.2018Turbulent,Moller.2022Combined}, this represents a significant limitation for their studies by numerical approaches. However, the large aspect ratio requires the design of novel experimental setups and the adaption and development of advanced temporally and spatially resolved velocity and temperature measurements. These are briefly discussed in \ref{Sec:Experimental methods}.  

\section{Methods}
\label{Sec:Methods}
\subsection{Experimental setup and measurements}
\label{Sec:Experimental methods}
\begin{figure}[tbp]
    \centering
    \includegraphics[width=0.46\textwidth]{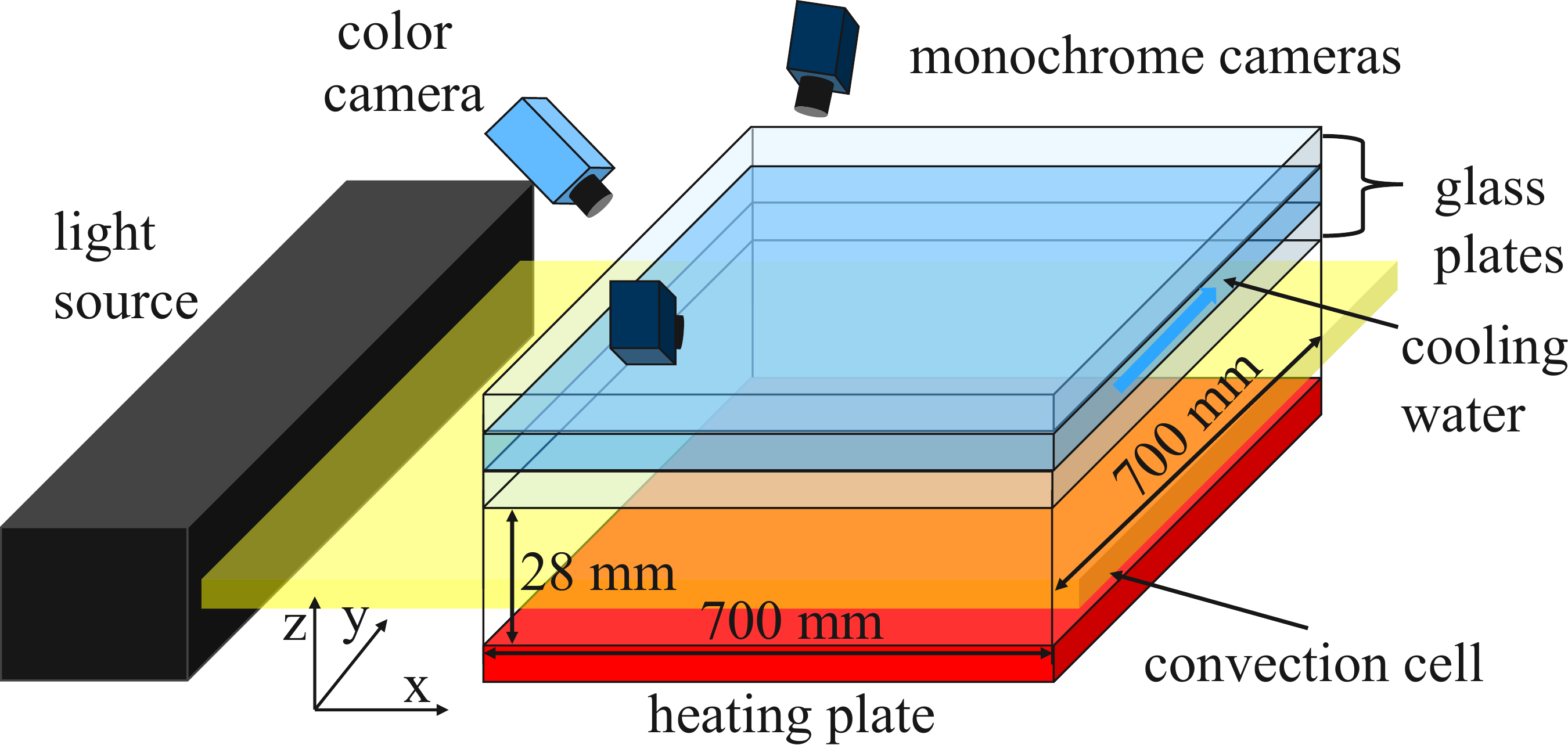}
       \centering
    \caption{Sketch of the experimental setup for the simultaneous measurements of all three velocity components and the temperature in a wide horizontal plane through a transparent cooling plate.}
    \label{fig:setup}
\end{figure}

All experiments that are considered in this report were conducted in a cuboid Rayleigh-Bénard convection cell with aspect ratio $\Gamma = 25$ and a width $W = 700$ mm and height $H = 28$ mm. The cell is filled with water at a mean temperature of $T_{\textrm{ref}} \approx 19.5^\circ \textrm{C}$ exhibiting a Prandtl number $\Pr = 7.1$. This cell design, with its large aspect ratio, allows the turbulent superstructures to form and is sketched in Figure \ref{fig:setup}. 
The fluid layer is confined by glass side walls, a heating plate at the bottom made of aluminum and a cooling plate assembly at the top made of two parallel glass sheets with a horizontal gap in between to allow cooling water to flow through. The temperature of both plates can be controlled independently  by thermostats.

Note here, in particular, that the transparent cooling plate arrangement is required to grant optical access to the flow domain from the top, and enables the large field of view (FOV) inevitable for the investigation of the large-scale flow structures or turbulent superstructures. Moreover, a comparably detailed study of the latter would not be possible in a vertical plane due to the small cell height~\cite{Cierpka.2019On}. To measure the fluid velocity and temperature simultaneously, a combined stereoscopic Particle Image Velocimetry (PIV) and Particle Image Thermometry (PIT) approach with  polymer-encapsulated thermochromic liquid crystals (TLCs) as seeding particles is employed. The unique property of a temperature-related wavelength change of the reflected light of the TLCs is leveraged to quantify the temperature. Hence, a light source with a continuous illumination spectrum is required since the TLCs can reflect only spectral components of light that are part of the original illumination spectrum. To achieve optimal illumination, a custom-made light source was designed, which consists of an LED array combined with light sheet optics to focus the light into a sheet of thickness $\approx$ 3-4 mm across the entire FOV.

Three cameras observe the flow domain with an observation angle of $\Phi \approx 65^\circ$. While two of those are monochrome cameras in a stereoscopic arrangement and used to perform stereoscopic PIV -- which measures all velocity components within a plane \cite{Prasad.2000Stereoscopic,Raffel.2018Particle} --, the third one is a color camera to simultaneously capture the color of the light that is reflected by the TLCs. Subsequently, the temperature in the ligth sheet can be inferred by means of PIT \cite{Dabiri.2009Digital}. In more detail, the color image is sliced into sub-images or interrogation windows, and the red, green and blue color values inside the interrogation windows are averaged. Those averages are subsequently transformed into the hue, saturation and value (HSV) color space. To determine the temperature, the hue is compared to a calibration curve. As the TLC reflection properties depend on the angle between the light source and observation, calibration curves are obtained during the calibration process for each interrogation window individually. Details on this technique can be found in Moller et al.~\cite{Moller.2019Influence}. Alternatively, a neural network can be trained to replace these local calibration functions \cite{Anders.2020Simultaneous,Moller.2020On}. The extension towards fully volumetric temperature and velocity measurements for wide observation regions, currently only available for low-aspact ratio setups~\cite{Schiepel_2021}, is still ongoing.

\begin{table*}[tb]
\centering
\begin{tabular}{cccccccc}
 Ra &  Ra$_{\mathrm{exact}}$& $\Nu$&$T_{\mathrm{h}}$ & $T_{\mathrm{c}}$ & $\tilde{x} \times \tilde{y}$ & $\tau_{\mathrm{f}}$ & $\tilde{t}_{\mathrm{total}}$ \\ \hline
 $2 \times 10^{5}$ &$2.07 \times 10^{5}$& 4.00 & $19.78^{\circ} \mathrm{C}$ & $19.08^{\circ} \mathrm{C}$ & $16.1 \times 16.7$ & $4.51 \mathrm{~s}$ & $5.39 \times 10^{3}$ \\
 $4 \times 10^{5}$ &$4.34 \times 10^{5}$& 5.31 & $20.22^{\circ} \mathrm{C}$ & $18.76^{\circ} \mathrm{C}$ & $16.1 \times 16.7$ & $3.12 \mathrm{~s}$ & $7.79 \times 10^{3}$ \\
 $7 \times 10^{5}$ &$7.34 \times 10^{5}$&5.58 & $20.87^{\circ} \mathrm{C}$ & $18.43^{\circ} \mathrm{C}$ & $16.2 \times 16.6$ & $2.40 \mathrm{~s}$ & $1.01 \times 10^{4}$  \\
\end{tabular}
\caption{Overview of the global parameters for the different experimental runs. Ra and Ra$_{\mathrm{exact}}$ denote the Ra we subsequently refer to and the exact Ra and Nu the measured global Nusselt number.  $T_{\mathrm{h}}$ and $T_{\mathrm{c}}$ represent the heating and cooling plate temperature, $\tilde{x}\times \tilde{y}$ the field of view, $\tau_{\mathrm{f}}$ the free-fall time and $\tilde{t}_{\textrm{total}}$ the duration of the experimental run in free-fall units. Data adapted from \cite{Moller.2022Experimental}.}
\label{tab:exp_details}
\end{table*}

In this paper, we consider three experimental runs at $\Ra = 2\times10^5$, $4\times10^5$, and $7\times10^5$ with measurements performed in the horizontal midplane. To reduce the amount of data and the required processing time, the temperature and velocity fields are measured in bursts of 5 minutes and inter-burst-breaks of 15 minutes. For each experiment, 19 bursts with 200 snapshots each are utilized. The dimensional data is subsequently non-dimensionalized as explained in section \ref{Sec:Numerical_methods} for the Dirichlet case and $\Ra \equiv \RaD$. Further details on the experimental runs are provided in Table \ref{tab:exp_details}.

\subsection{Numerical simulations}
\label{Sec:Numerical_methods}
To compare the experimental setup with numerically obtained results, we consider the simplest considerable turbulent convection configuration -- three-dimensional, turbulent Rayleigh-Bénard convection in the Oberbeck-Boussinesq approximation. In this setup, buoyancy drives the flow by coupling the scalar temperature field $T \left( \bm{x}, t\right)$ with the incompressible velocity vector field $\bm{u} \left( \bm{x}, t \right)$ where $\bm{x} = (x, y, z)$ and $\bm{u} = (u_{x}, u_{y}, u_{z})$. In more detail, the mass density varies in this approximation only in the buoyancy forcing term, and there only at first order with the temperature deviation from some reference value.

Similar to the experimental setup, we consider a Cartesian domain $V = W \times W \times H$. The confining bottom and top planes at $z \in \left\{ 0, H \right\}$ obey either no-slip (ns) or free-slip (fs) mechanical boundary conditions (mbcs),
\begin{equation}
    u_{x} = u_{y} = u_{z} = 0 \qquad \textrm{or} \qquad u_{z} = \frac{\partial u_{x}}{\partial z} = \frac{\partial u_{y}}{\partial z} = 0 ,
\end{equation}
respectively. The lateral boundaries (lbs) are either closed and insulating -- in the case of which they exhibit again no-slip boundary conditions --, or periodic. In the case of the latter, any quantity $\Phi$ is repeated after the periodic length $W=W_x=W_y$ such that
\begin{equation}
    \Phi \left( \bm{x}, t \right) = \Phi \left( \bm{x} + i_{x} W_{x} \bm{e}_{x} + i_{y} W_{y} \bm{e}_{y}, t \right) \quad \textrm{with } i \in \mathbb{Z} .
\end{equation}

In accordance with the scientific objective of the present study, we vary the thermal boundary conditions (tbcs) at the plates to correspond mathematically either to a Dirichlet (D) or a Neumann (N) condition. In the case of Dirichlet conditions, the top and bottom planes exhibit \textit{constant temperatures}
\begin{equation}
    T \left( z = 0 \right) = T_{\textrm{bot}} \quad \textrm{and} \quad T \left( z = H \right) = T_{\textrm{top}} ,
\end{equation}
whereas in the Neumann case the temperature is allowed to vary locally -- all that is fixed is the applied constant heat flux through the spatially \textit{constant temperature gradient}
\begin{equation}
    \left. \frac{\partial T}{\partial z} \right|_{z = 0} = \left. \frac{\partial T}{\partial z} \right|_{z = H} = - \beta \quad \textrm{with } \beta > 0 .
\end{equation}

The equations of motion are made dimensionless based on characteristic quantities of the system. We adopt the layer height $H$ as unit of length and the so-called free-fall time $\tau_{\textrm{f}} = H / U_{\textrm{f}}$ as the unit of time with the free-fall velocity $U_{\textrm{f}} = \sqrt{\alpha g T_{\textrm{char}} H}$. Here, $\alpha$ represents the isobaric expansion coefficient and $g$ the acceleration due to gravity. $T_{\textrm{char}}$ is a characteristic temperature which depends on the thermal boundary condition. In the Dirichlet case $T_{\textrm{char}} = \Delta T = T_{\textrm{bot}} - T_{\textrm{top}} > 0$, whereas in the Neumann case $T_{\textrm{char}} = \beta H$.
The governing equations follow in their non-dimensional description as
\begin{align}
\label{eq:CE}
\tilde{\nabla} \cdot \tilde{\bm{u}} &= 0 , \\
\label{eq:NSE}
\frac{\partial \tilde{\bm{u}}}{\partial \tilde{t}} + ( \tilde{\bm{u}} \cdot \tilde{\nabla} ) \thickspace \tilde{\bm{u}} &= - \tilde{\nabla} \tilde{p} + \sqrt{\frac{\Pr}{\Ra_{\textrm{D,N}}}} \tilde{\nabla}^{2} \tilde{\bm{u}} + \tilde{T} \bm{e}_{z} , \\
\label{eq:EE}
\frac{\partial \tilde{T}}{\partial \tilde{t}} + ( \tilde{\bm{u}} \cdot \tilde{\nabla} ) \thickspace \tilde{T} &= \frac{1}{\sqrt{\Ra_{\textrm{D,N}} \Pr}} \tilde{\nabla}^{2} \tilde{T} ,
\end{align}
where $p$ is the pressure field, and tildes indicate dimensionless quantities. Two dimensionless parameters -- the Prandtl number \Pr and the Rayleigh number \Ra -- control the entire dynamics and summarise all the dimensional coefficients and parameters. These numbers are defined as
\begin{equation}
\label{eq:def_Pr_RaD_RaN}
\Pr = \frac{\nu}{\kappa} , \thickspace \thickspace \thickspace
\RaD = \frac{g \alpha \Delta T H^3}{\nu \kappa} , \textrm{ and} \thickspace \thickspace \thickspace
\RaN = \frac{g \alpha \beta H^4}{\nu \kappa} 
\end{equation}
with the kinematic viscosity and temperature diffusivity of the fluid, $\nu$ and $\kappa$, respectively. Note that due to the non-dimensionalization, e.g., the vertical coordinate $0 \leq \tilde{z} \leq 1$. 

We solve the equations of motion using the spectral element solver \textsc{nek5000} \citep{Fischer1997,Scheel2013} and resolve all dynamically relevant scales down to the Kolmogorov scale $\eta_{\textrm{K}}$. All simulations start with the fluid at rest and the linear conduction temperature profile, which is randomly and infinitesimally perturbed. The global mean temperature is thus independent of the thermal boundary condition, $\langle \tilde{T} \rangle_{\tilde{V}} = 0.5$. More detailed information on numerical details can be found in our previous works \citep{Scheel2013, Pandey.2018Turbulent, Vieweg.2021Supergranule, Vieweg.2021Lagrangian,fonda2019deep}.

Finally, the global turbulent heat transfer across the fluid layer is quantified by the Nusselt number $\Nu$, which is given for the Dirichlet case by 
\begin{equation}
\label{eq:def_NuD}
    \NuD = - \Bigg\langle \left. \frac{\partial \tilde{T}}{\partial \tilde{z}} \right|_{\tilde{z} = 0} \Bigg\rangle_{\tilde{A}} = - \Bigg\langle \left. \frac{\partial \tilde{T}}{\partial \tilde{z}} \right|_{\tilde{z} = 1} \Bigg\rangle_{\tilde{A}} 
\end{equation}
where $\tilde{A}$ is the non-dimensional horizontal cross section $\tilde{A} = \Gamma \times \Gamma$, and for the Neumann case by
\begin{equation}
\label{eq:def_NuN}
    \NuN = \frac{1}{\langle \tilde{T} \left( \tilde{z} = 0 \right) \rangle_{\tilde{A}} - \langle \tilde{T} \left( \tilde{z} = 1 \right) \rangle_{\tilde{A}}} .
\end{equation}
This allows eventually to connect the different definitions of the Rayleigh numbers via $\RaD = \RaN /\NuN$ \citep{Otero2002}.
In total, we consider three numerical data sets with various boundary conditions -- details are summarised in Table \ref{tab:num_details}
\begin{table}[tb]
\centering
\begin{tabular}{ccccccc}
 Ra & Pr& $\Gamma$& mbc & tbc & lb & $\tilde{\lambda}$ \\ \hline
 $1 \times 10^{5}$ & 7 & 25 & ns & D & closed & 6.26  \\
 $1 \times 10^{6}$ & 7& 25 & ns & D & closed & 6.26   \\
 $2 \times 10^{5}$ & 7 & 60 & fs & N & periodic & $\leq$ 60    \\
\end{tabular}
\caption{Table of the parameters of the different numerical runs. Including the Rayleigh number \Ra, the Prandtl number \Pr, the aspect ratio $\Gamma$, the mechanical and thermal boundary condition, the lateral boundaries and the characteristic large-scale flow structure size $\tilde{\lambda}$.}
\label{tab:num_details}
\end{table}

\section{The influence of the thermal boundary conditions on the large-scale flow structure size}
\label{Sec:superstructures}
As highlighted in the introduction, one of the disagreements observed by Moller et al.~\cite{Moller.2022Combined} concerned the size of the large-scale flow structures -- in particular, the experimental data showed larger structures compared to DNS data with isothermal boundary conditions at similar $\Ra$. The authors attributed this disagreement to the thermal boundary conditions at the cooling plate. Due to the low thermal conductivity of glass the cooling plate cannot be considered as perfectly isothermal.

\begin{figure}[tb]
    \centering
    \includegraphics[width=\linewidth]{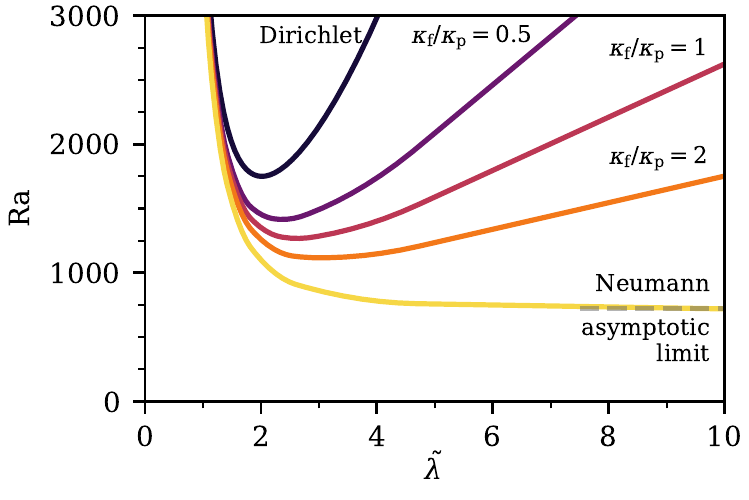}
       \centering
    \caption{Linear stability at varying thermal but mechanical no-slip boundary conditions. As the ratio of thermal diffusivities deviates stronger from the Dirichlet scenario, the critical structures at the onset of convection become less stable and increasingly extended. The Dirichlet case is computed according to \citep{Bestehorn2006}, whereas the other curves are interpolated from \citep{Hurle1967}.}
    \label{fig:lin_stability}
\end{figure}

Such a connection between the convective structure size and thermal boundary condition is actually already supported by the linear stability at the onset of convection. Figure \ref{fig:lin_stability} visualizes the neutral stability curves for various thermal boundary conditions. While the Dirichlet case with $\kappa_{\textrm{f}} / \kappa_{\textrm{p}} \rightarrow 0$ is the most stable one, the opposing Neumann case in the limit $\kappa_{\textrm{f}} / \kappa_{\textrm{p}} \rightarrow \infty$ is the least stable one. Between these two extreme cases, many other neutral curves can be drawn depending on the ratio of thermal diffusivities of the fluid and the solid plate $\kappa_{\mathrm{f}} / \kappa_{\mathrm{p}}$. Interestingly, even the critical convective structure size seems to change gradually from one limit to the other, see Fig. \ref{fig:lin_stability}. 

As convection sets in, heat conduction is complemented by convective heat transfer. While the latter's strength is quantified by the Nusselt number $\Nu$, a meaningful amendment to the ratio of thermal diffusivities is provided by the effective ratio $\kappa_{\textrm{f}} \Nu / \kappa_{\textrm{p}}$. Physically, the latter controls the propagation or relaxation of thermal perturbations \citep{Hurle1967}. An alternative measure to categorize the thermal boundary condition is the Biot number $\Bi$, which describes the ratio of convective heat transfer at the surface of a body to the heat transfer inside the body due to pure heat conduction \cite{polifke2009warmeubertragung}. Translated to our experimental setup, this quantity is given by
\begin{equation}
\Bi = \Nu \frac{\lambda_{\mathrm{f}} d_{\mathrm{p}}}{\lambda_{\mathrm{p}} H} 
\end{equation}
with the plate thickness $d_{\mathrm{p}}$. 

\begin{figure*}[tb]
    \centering
    \includegraphics[width=\textwidth]{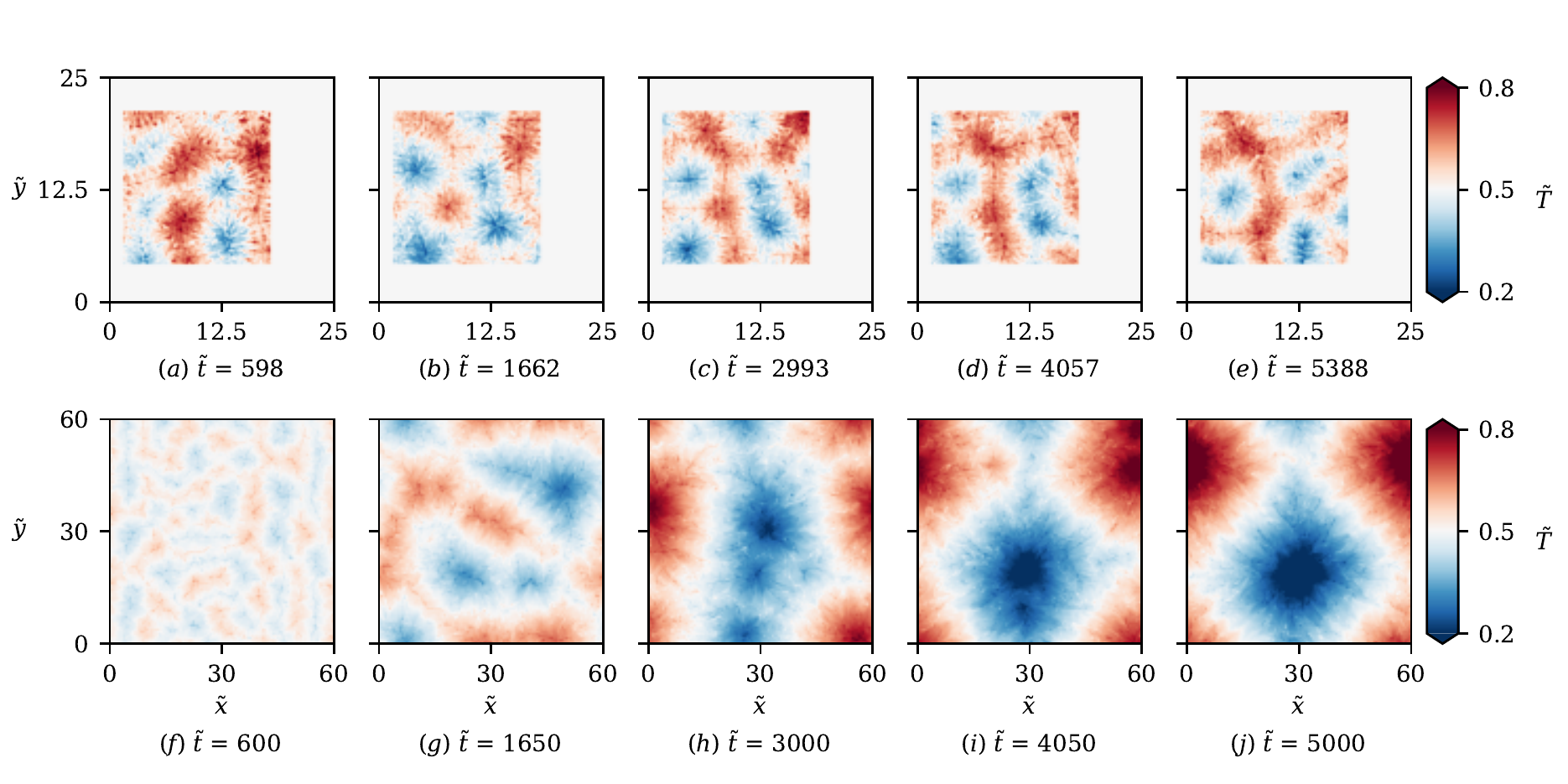}
       \centering
    \caption{Temperature field in the horizontal midplane obtained from measurements at \Ra = $2\times10^{5}$ (top row) and from simulations with thermal Neumann boundary conditions and mechanical free-slip boundary conditions at $\RaN = 2 \times 10^{5}$ (bottom row) at five exemplary time instances. 
    To remove noise induced by the measurements, the experimental temperature fields are time-averaged over $33 \tau_{\textrm{f}}$.}
    \label{fig:temporal_development_fields}
\end{figure*}

Differently to simulations, perfect thermal boundary conditions cannot be achieved in any experiments. However, when the ratio of the effective thermal diffusivities $\kappa_{\textrm{f}} \Nu / \kappa_{\textrm{p}} \ll 1$ and $\Bi \ll 1$, isothermal conditions can be well approximated. We summarise the corresponding ratios or values for all experiments presented in this paper in Table \ref{tab:boundary_conditions}. As can be seen, both characteristics depend crucially on the choice of the plate. 
On the one hand, the heating plate can be considered isothermal since $\Bi_\mathrm{h}$ and $\kappa_{\textrm{f}} \Nu / \kappa_{\textrm{p, h}}$ are both of the order $\mathcal{O} \left( 10^{-2} \right)$.
On the other hand, at the cooling plate, both $0.88 \lesssim \Bi_\mathrm{c} \lesssim 1.22$ and $1.47 \lesssim \kappa_{\textrm{f}} \Nu / \kappa_{\textrm{p, c}} \lesssim 2.05$ -- depending on the particular Rayleigh number -- indicate that the thermal boundary conditions can neither be considered as constant temperature nor constant heat flux, but instead only as mixed boundary condition. Recalling the stability behavior at the onset of convection -- see again Figure \ref{fig:lin_stability} --, these numbers also imply larger most unstable convective length scales.
\begin{table}[tb]
    \centering
    \begin{tabular}{c @{\hskip 2mm}  c @{\hskip 2mm} c @{\hskip 2mm} c @{\hskip 2mm} c @{\hskip 2mm} c @{\hskip 2mm} c}
        $\Ra$  & $\Bi_\mathrm{c}$ & $\Bi_\mathrm{h}$ & $\kappa_{\mathrm{f}} \Nu/ \kappa_{\textrm{p, c}}$&  $\kappa_\mathrm{f} \Nu/ \kappa_\mathrm{\textrm{p, h}}$ & $\tilde{\lambda}$ \\ \hline
        $2 \times 10^{5}$  & 0.88 & 0.0091 & 1.47 & 0.0092 & 9.36  \\ 
        $4 \times 10^{5}$  & 1.16 & 0.0121 & 1.95 & 0.0121 & 10.53  \\ 
        $7 \times 10^{5}$  & 1.22 & 0.0127 & 2.05 & 0.0128 & 11.50  \\ 
    \end{tabular}
    \caption{Table of the Biot number at the cooling and heating plate Bi$_\mathrm{c}$ and Bi$_\mathrm{h}$, the ratio of the effective thermal diffusivities of the fluid and the plate material at the cooling and heating plate $\kappa_\mathrm{f} \Nu/ \kappa_\mathrm{p, c}$ and $\kappa_\mathrm{f} \Nu/ \kappa_\mathrm{p, h}$, and the average characteristic size of the temperature patterns $\tilde{\lambda}$ for the experiments at varying $\Ra$.}
    \label{tab:boundary_conditions}
\end{table}

Recent numerical studies showed that the underlying stability from at and slightly above the onset of convection is not forgotten by the dynamical system even when the latter is driven far into the turbulent regime \cite{Vieweg.2021Supergranule, Vieweg.2022Inverse}. Transferring this consideration to the present context allows us to identify the top plate as an important factor influencing the formation of large-scale flow structures. In other words, the primary stability behavior suggests the formation of larger convective flow patterns even in the turbulent regime, particularly for the top plate. This agrees, in fact, with our experimental observations. As the Rayleigh number $\Ra$ is increased, so is the effective ratio due to the larger Nusselt number, and the experimentally observed patterns become successively larger -- see again Table \ref{tab:boundary_conditions}.

Clearly, the size of the large-scale flow structures manifests based on a balance between almost perfect Dirichlet boundary conditions at the bottom plate and conditions closer to the Neumann case at the top plate. A significantly height-dependent flow structure size should, however, not be expected as $\tilde{\lambda} \gg H$.

\begin{figure*}[tb]
    \centering
    \includegraphics[width=1\textwidth]{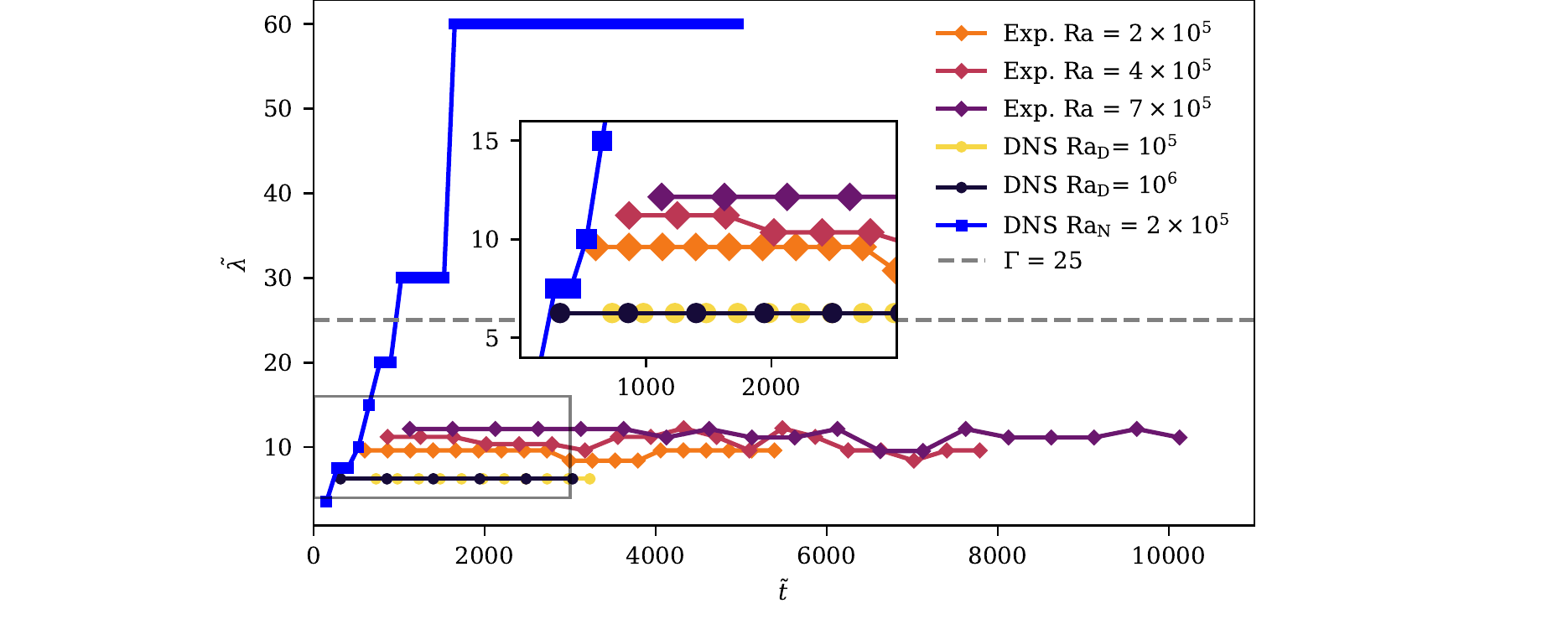}
       \centering
    \caption{Plot of the temporal development of the superstructure size. The horizontal axis shows the time in free-fall units $\tau_{\textrm{f}}$, and the vertical axis shows the structure size. The dashed line indicates an aspect ratio $\Gamma$=25, which is the domain size of the experiment and the numerical simulation in the Dirichlet configuration. 
    }
    \label{fig:structure_size_over_time}
\end{figure*}

As outlined in Section \ref{Sec:turb_superstructures}, different thermal boundary conditions result furthermore in different temporal behavior of the large-scale flow structures. We compare this aspect in Figure \ref{fig:temporal_development_fields} for numerical and experimental data. Note that the experimental data visualize an over $33 \tau_{\textrm{f}}$ slightly time-averaged temperature field (to reduce the influence of noise on the shape of the superstructures), whereas the numerical data shows instantaneous snapshots throughout the gradual supergranule aggregation for the Neumann case. The temporal behavior of both flows differs significantly. While the structures observed in the experiment show a rather slow reorganization over time, the numerical Neumann case shows an intense gradual aggregation process. The fundamental difference concerning the long-term behavior or dynamics of the large-scale flow structures suggests to term the experimentally observed patterns are still \textit{turbulent superstructures} rather than \textit{supergranules}.

This is supported by an even more comprehensive comparison of the temporal evolution of the large-scale flow structure size for all in this work considered experimental and numerical data. We compute, therefore, the two-dimensional power spectrum of the temperature field, average it in azimuthal direction, determine the wave number $\hat{k}$ of the spectral peak, and eventually calculate the corresponding characteristic wavelength via $\tilde{\lambda} = 2 \pi / \hat{k}$. Note here that, as already above, the experimental temperature fields are again averaged over $33 \tau_{\textrm{f}}$ and the two-dimensional spectra zero-padded to reduce the influence of noise and increase the spectral resolution, respectively. The results of this analysis are summarized in Figure \ref{fig:structure_size_over_time}. Note here that the discrete nature of this evolution is nothing but the consequence of the discrete wave numbers corresponding to the Discrete Fourier Transform (DFT). Here, we want to point out that the mentioned structure size appears larger, as reported in previous publications. This is exclusively related to a different binning scheme of the azimuthally averaged spectra required to capture the structure size in the Neumann case correctly. When looking at this Figure from the bottom to the top, the turbulent superstructures observed in the constant temperature simulations exhibit no variation in the structure size over time. The situation changes for the experimental runs, in which the structure size is significantly larger and fluctuates slightly over time. This fluctuation can be related to the rearrangement of the turbulent superstructures and the horizontally not entirely captured flow domain. Unlike the Dirichlet simulations, the structure size in the experiments clearly depends on the Rayleigh number $\Ra$. We attribute this circumstance to the $\Ra$-dependent thermal boundary conditions at the top plate, as explained above. Although the structure size fluctuates over time in the experiment, this is in clear contrast to the case of thermal Neumann boundary conditions that are applied at both plates in the remaining numerical simulation. Here, the gradual aggregation of the flow structures results in a temporal evolution that stops only artificially once the domain size of $\Gamma = 60$ is reached.

These comparisons of the temporal evolution of the large-scale flow structures underline the different natures of flow structures subject to different thermal boundary conditions. In a nutshell, the experimentally observed patterns are unambiguously more similar to the turbulent superstructures known for Dirichlet conditions than to those found for Neumann conditions. Again, this is in accordance with the still finite critical convective structure size at the onset of convection for $\kappa_{\textrm{f}} / \kappa_{\textrm{p}} = 2$, recall Figure \ref{fig:lin_stability}.

\section{The influence of the thermal boundary conditions on the heat transfer}
\label{Sec:heat_transfer}
Likewise to the difference in structure size, Moller et al. \cite{Moller.2022Combined} reported differences in the heat transfer between their experiments and numerical simulations. In particular, the former showed a reduced global heat transport as well as higher probabilities of negative local Nusselt numbers \Nuloc (which will be defined in eq. \eqref{eq:Nu_loc_Theta}) compared to the latter. First, we want to understand in more detail how the global and local Nusselt number is altered by the experimental measurements and its definition, respectively. 

We start by defining the local Nusselt number $\Nu$ similar to \cite{Vieweg.2021Lagrangian}
\begin{equation}
\NulocT \left( \tilde{\bm{x}}, \tilde{t} \right) = \sqrt{\Ra\Pr} \thickspace \tilde{u}_z \left( \tilde{\bm{x}}, \tilde{t} \right) \tilde{T} \left( \tilde{\bm{x}}, \tilde{t} \right) - \frac{\partial \tilde{T} \left( \tilde{\bm{x}}, \tilde{t} \right)}{\partial \tilde{z}}
\label{eq:Nu_loc_T}
\end{equation}
which is proportional to the vertical component of the heat current vector.
Note that although both \NulocT and \Nuloc are local measures, the latter is a subset of the former. This is done by decomposing the temperature into the linear conduction profile $\tilde{T}_{\mathrm{lin}}$ and its fluctuating part or deviation $\tilde{\Theta}$ such that $\tilde{T} \left( \tilde{\bm{x}}, \tilde{t} \right) = \tilde{T}_{\mathrm{lin}} \left( \tilde{z} \right) + \tilde{\Theta} \left( \tilde{\bm{x}}, \tilde{t} \right)$ with $\tilde{T}_{\mathrm{lin}} = 1 - \tilde{z}$. Hence, the local Nusselt number is the combination of $4$ separate subsets such that $\NulocT = \sum_{i = 1}^{4} \Nu_{\textrm{loc,}i}$ with
\begin{align}
    \Nu_{\textrm{loc,}1} &:= \sqrt{\Ra\Pr} \thickspace \tilde{u}_z\left( \tilde{\bm{x}}, \tilde{t} \right) \tilde{T}_{\textrm{lin}} ,\\
    \label{eq:Nu_loc_Theta}
    \Nu_{\textrm{loc,}2} &:= \sqrt{\Ra\Pr} \thickspace \tilde{u}_z\left( \tilde{\bm{x}}, \tilde{t} \right) \tilde{\Theta}\left( \tilde{\bm{x}}, \tilde{t} \right) \equiv \Nuloc , \\
    \Nu_{\textrm{loc,}3} &:= - \frac{\partial \tilde{T}_{\textrm{lin}}}{\partial \tilde{z}} = 1 ,\\
    \Nu_{\textrm{loc,}4} &:= - \frac{\partial \tilde{\Theta}\left( \tilde{\bm{x}}, \tilde{t} \right)}{\partial \tilde{z}} .
\end{align}
Here we dropped the designation of dependence of the local Nusselt numbers on $\left( \tilde{\bm{x}}, \tilde{t} \right)$  for the sake of conciseness.
To justify a restriction on $\Nuloc$, one may consider (i) horizontal averages across the entire cross-section and (ii) an analysis at midplane $z = 0.5$. In this case, $\langle \tilde{u}_{z} \rangle_{\tilde{A}} = 0$ due to continuity and $\langle \partial \tilde{\Theta} / \partial \tilde{z} \left( \tilde{z} = 0.5 \right) \rangle_{\tilde{A}} = 1$, so $\langle \Nu_{\textrm{loc,}1} \rangle_{\tilde{A}} = 0$ and $\langle \Nu_{\textrm{loc,}3} \rangle_{\tilde{A}} + \langle \Nu_{\textrm{loc},4} \rangle_{\tilde{A}} = 0$ follows at midplane, respectively.
Given these considerations, one ends up with 
\begin{equation}
\label{eq:matching_local_Nu_definitions}
\langle \NulocT \left( \tilde{z} = 0.5 \right) \rangle_{\tilde{A}} = \langle \Nuloc \left( \tilde{z} = 0.5 \right) \rangle_{\tilde{A}} .
\end{equation}
Note that these Nusselt numbers yield the global values of the heat transfer, so $\Nu = \langle \Nuloc \left( \tilde{z} = 0.5 \right) \rangle_{\tilde{A}}$.

Clearly, the experimental measurements do not meet the two assumptions or considerations from above perfectly due to the limited field of view and the light sheet thickness, respectively. Besides the circumstance of not being able to measure the vertical gradient of the temperature field, the statistic of $\Nuloc$ in the experiment is further altered by a coarser measurement resolution. In more detail, the experimental measurements cannot resolve the Kolmogorov scale, which in turn affects the statistics. Thus, there are effects of the experimental measurement on both the \textit{global} and \textit{local} Nusselt number.

\begin{figure}[tb]
    \centering
    \includegraphics[width=1\linewidth]{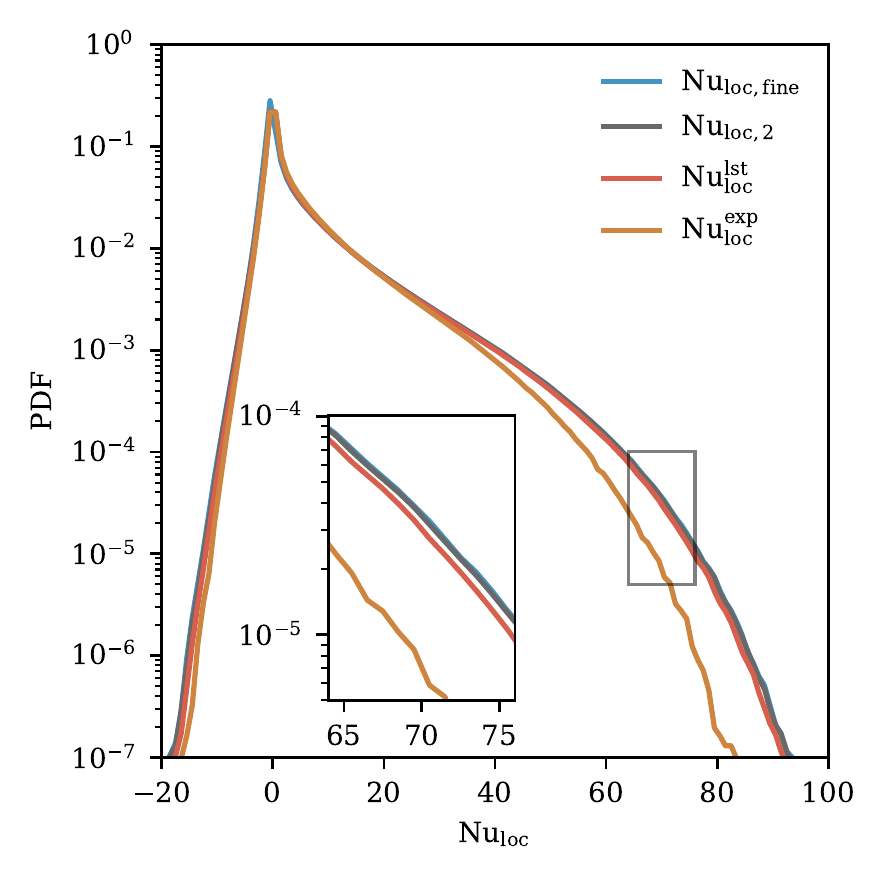}
       \centering
    \caption{Local Nusselt number analysis at or around the horizontal midplane for the numerical data at \RaD= 1$\times10^5$.}
    \label{fig:Pdf_influence}
\end{figure}
In order to investigate the impact of these experimental effects on the statistics, we compute the probability density functions (PDFs) of descendants of $\NulocT$ for the well-resolved numerical data at $\Ra = 1 \times 10^{5}$ with different pre-processing. For a more concise representation, we omit here the effect of $\Nu_{\textrm{loc,}1}$ as its statistic is symmetric around zero if being evaluated across the entire cross-section. We start with $\Nu_{\textrm{loc, fine}}$, which represents the PDF of $\NulocT - \Nu_{\textrm{loc,}1}$ based on high-resolution data at $z = 0.5$.  Starting from this quantity, we additionally neglect the effect of heat conduction via $\Nu_{\textrm{loc,}3}$ and $\Nu_{\textrm{loc,}4}$, so only $\Nu_{\textrm{loc,}2}$ remains on the high-resolution grid. 

In the next step, we add the effect of the light sheet thickness by vertically averaging $\tilde{u}_{z}$ and $\tilde{\Theta}$ over the light sheet thickness $\approx 0.14 H$ centered around the midplane. We designate the resulting local Nusselt number as $\Nu_{\textrm{loc}}^{\textrm{lst}}$. As a final step and in addition to the averaging over the light sheet thickness, we perform an in-plane binning from the high resolution down to a spatial resolution that matches experimental conditions -- we denote this final quantity by $\Nulocexp$. Figure \ref{fig:Pdf_influence} summarises the PDFs for these four introduced quantities.

We find that the PDFs collapse quite well as long as no horizontal binning is performed. On the one hand, this underlines that heat conduction is not important in the bulk region, while, on the other hand, it judges the influence of the light sheet thickness as negligible. In contrast, in-plane binning shows the largest influence on the statistics. In essence, this averaging leads to an underestimation of the probability of large-magnitude events -- thus, this effect concerns the positive $\Nuloc$ range more than its negative counterpart. As the effect is not overwhelmingly strong, this underlines that the measurements are well suited to determine $\Nuloc \equiv \Nu_{\textrm{loc,}2} \approx \Nulocexp$ experimentally. 

\begin{figure}[tb]
    \centering
    \includegraphics[width=1\linewidth]{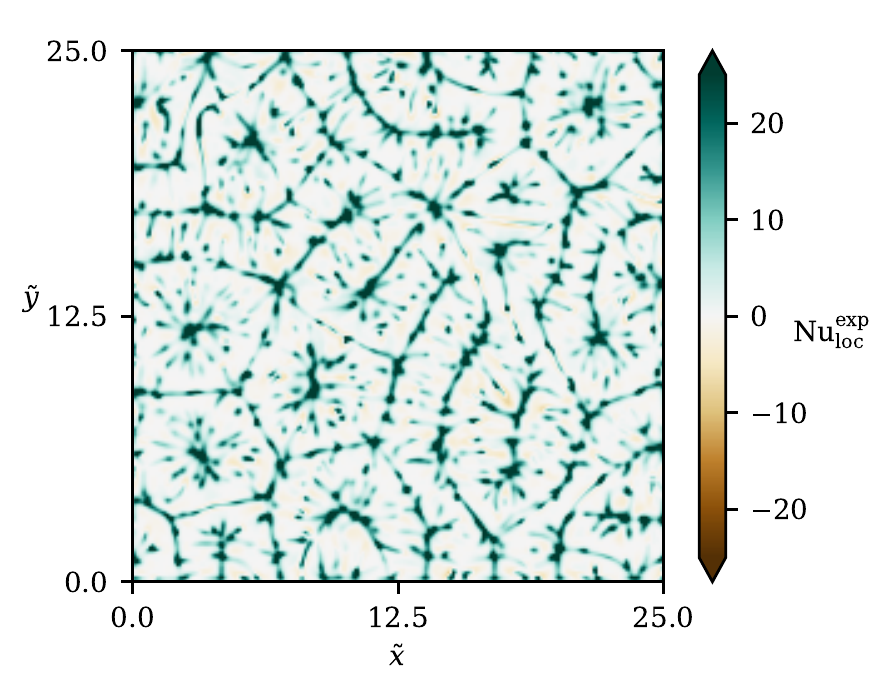}
       \centering
    \caption{Instantaneous $\Nulocexp$ field at the horizontal midplane. The data corresponds to $\RaD = 1 \times 10^5$.}
    \label{fig:Nu_fields_DNS}
\end{figure}
Yet, this sparser in-plane resolution must be taken into account when comparing experimental and numerical data. An exemplary snapshot of the resulting $\Nulocexp$ is shown in Figure \ref{fig:Nu_fields_DNS} for numerically generated data. As can be seen, the superstructures can -- together with their increased heat transfer at the ridges, see also Fonda et al.~\cite{fonda2019deep} -- easily be recognized despite the spatial binning. Note that deviations caused by this binning mostly affect the large-magnitude ridges but can practically not be recognized in such a plot. Hence, the binning does not distort the general picture of the superstructures and may also be applied in our subsequent data analysis. 

\begin{figure}[tb]
    \centering
    \includegraphics[width=\linewidth]{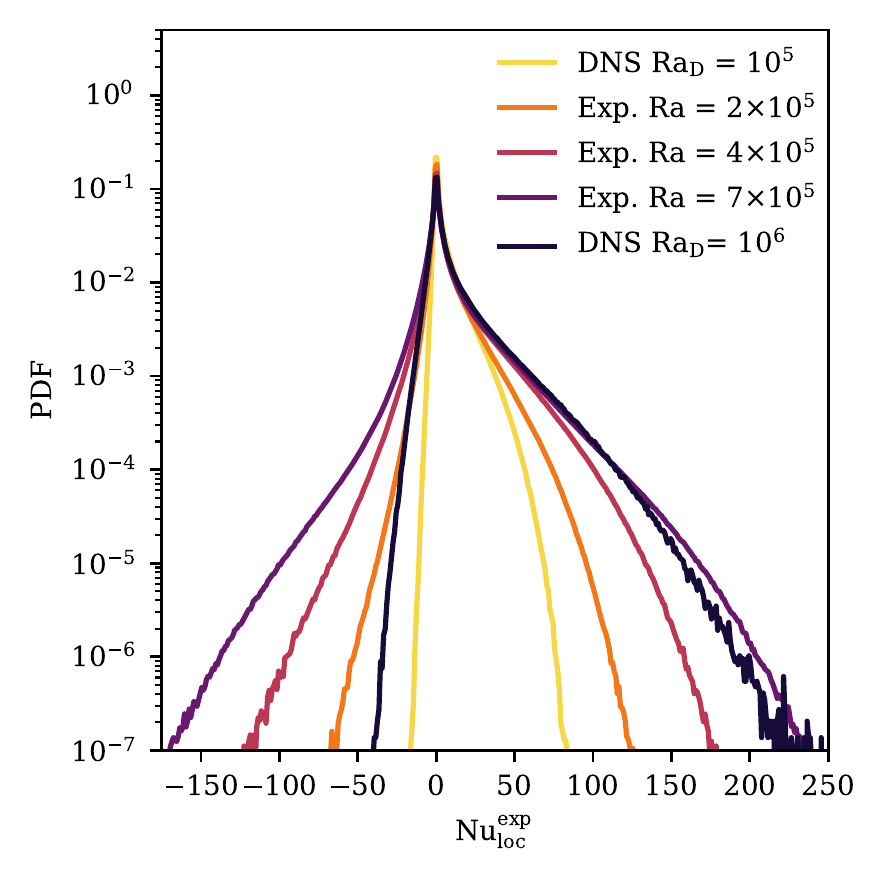}
       \centering
    \caption{PDF of the \Nulocexp for the experiments and the Dirichlet DNS.}
    \label{fig:Nu_exp_sim_comp_pdf}
\end{figure}
We contrast $\Nulocexp$ statistics from experimental and numerical data in Figure \ref{fig:Nu_exp_sim_comp_pdf}, indicating two interesting aspects. 
First, we observe a qualitatively reasonable trend with increasing $\Ra$ up to $\Ra = 7 \times 10^{5}$ for the \textit{positive} tails. Afterward, the trend seems to be broken, and the data contradicting. However, in our view, the data is still plausible since experimental measurements are subject to measurement uncertainties -- their PDFs spread unavoidably across broader ranges, leading to relatively higher probabilities of far-tail events. Thus, we conclude the trend to be quite consistent across the entire range of investigated $\Ra$. This first aspect is in clear contrast to the second one, the latter of which represents a clear mismatch for the \textit{negative} tails. More precisely, we observe a striking trend towards more frequent and extreme negative $\Nuloc$ events with increasing $\Ra$. This overshooting cannot be explained by the weak effect of measurement uncertainties from above, asking clearly for a more detailed analysis in the following.

\begin{figure*}[htbp]
    \centering
    \includegraphics[width=\linewidth]{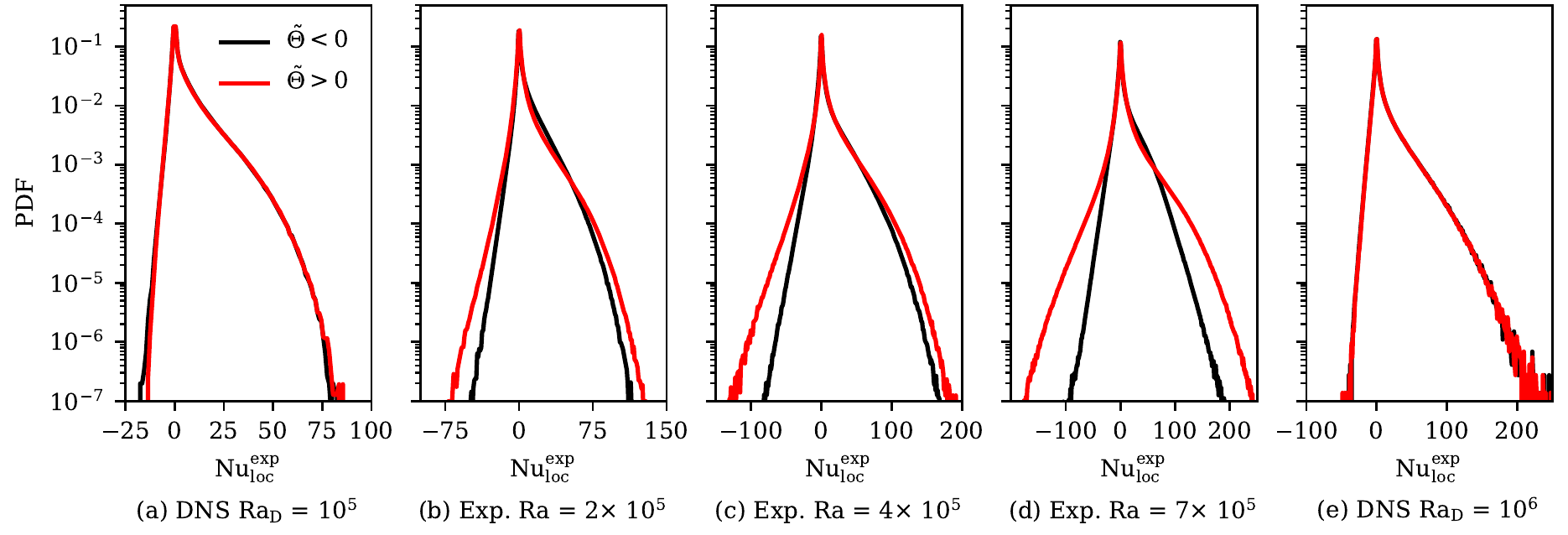}
       \centering
    \caption{PDFs of the decomposed $\Nulocexp$. The black and red lines correspond to contribution to $\Nulocexp$ caused by $\tilde{\Theta}<0$ and $\tilde{\Theta}_z>0$, respectively.
    }
    \label{fig:signed_pdf_experiment}
\end{figure*}
\begin{figure}[tb]
    \centering
    \includegraphics[width=0.47\textwidth]{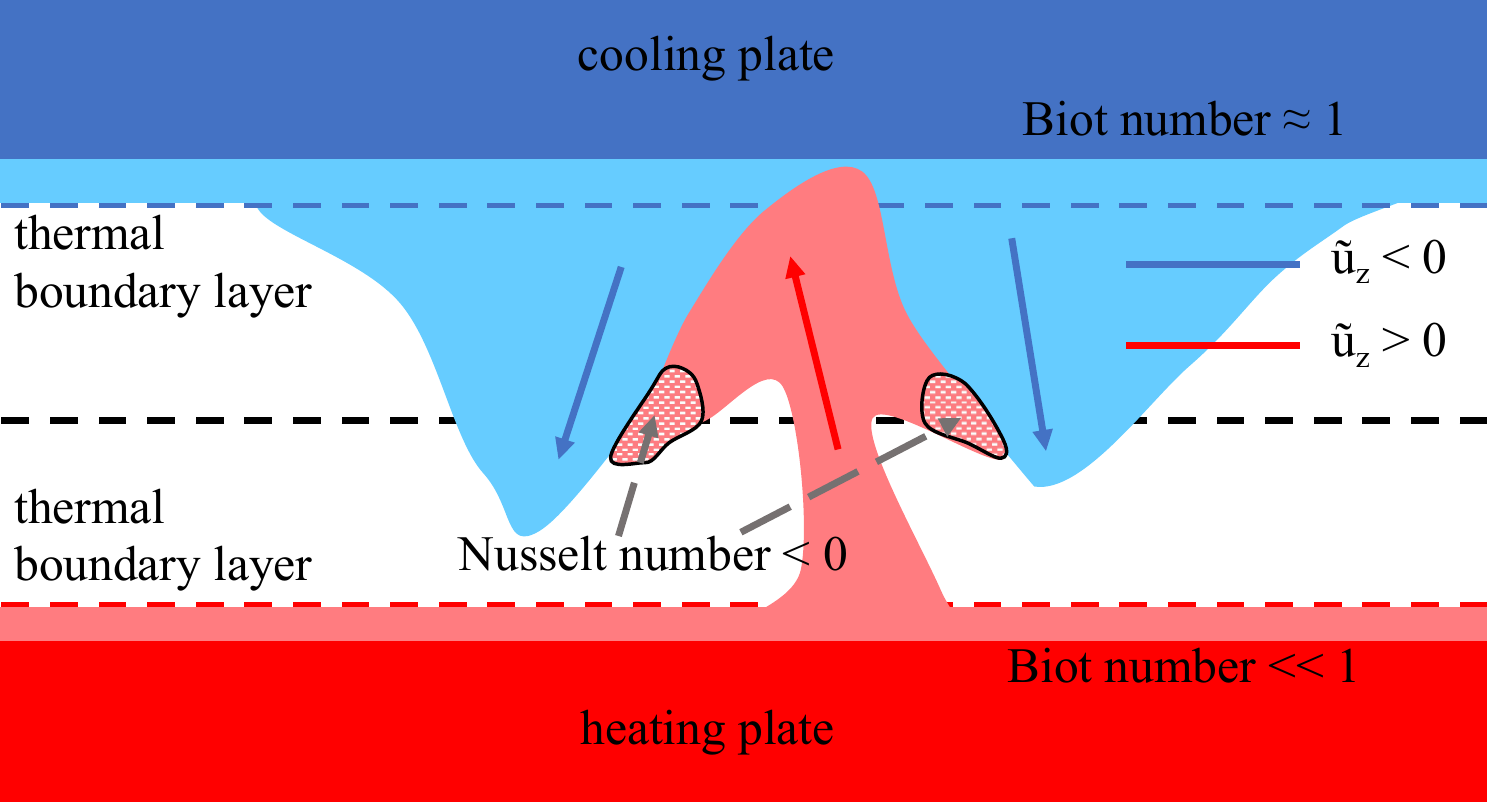}
       \centering
    \caption{Sketch of a negative $\Nuloc$ event. A hot thermal plume rises towards the cooling plate. Due to the lower thermal conductivity of the cooling plate, the heat of the hot thermal plumes is only slowly transferred through the plate. Meanwhile, detaching cold thermal plumes drag warm fluid with them, leading to negative $\Nuloc$ events.}
    \label{fig:sketch_reverse_flow}
\end{figure}

Recalling eq. \eqref{eq:Nu_loc_Theta} (to which $\Nulocexp$ is directly related), negative $\Nulocexp$ can either be caused by up-welling cold fluid or by down-welling hot fluid -- thus, we disentangle $\Nulocexp$ subsequently based on the sign of the temperature deviation field. We contrast the resulting conditioned PDFs of $\Nulocexp|_{\tilde{\Theta}<0}$ (black) and $\Nulocexp|_{\tilde{\Theta}>0}$ (red) for the entire range of $\Ra$ in Figure \ref{fig:signed_pdf_experiment}. First, we notice a perfect match of both PDFs for the numerically obtained data due to the top-down symmetry as a consequence of the underlying Oberbeck-Boussinesq approximation. In contrast, we find an increasing probability of (negative and positive) far-tail events in the case of hot fluid for experimental data and successively larger $\Ra$. This observation confirms an incapability of the cooling plate to generate comparable localized thermal variance -- that leads to high-magnitude $\Nulocexp$ events -- due to the smaller thermal conductivity of the glass plate. In other words, the cooling plate cannot transfer the heat out of up-welling thermal plumes fast enough. As a consequence, hot fluid that rose upwards is not sufficiently cooled in a region close to the top plate and is subsequently pulled down again by down-welling thermal plumes due to viscosity. Figure \ref{fig:sketch_reverse_flow} visualizes this concept in a sketch.

\begin{figure*}[tb]
    \centering
    \includegraphics[width=\textwidth]{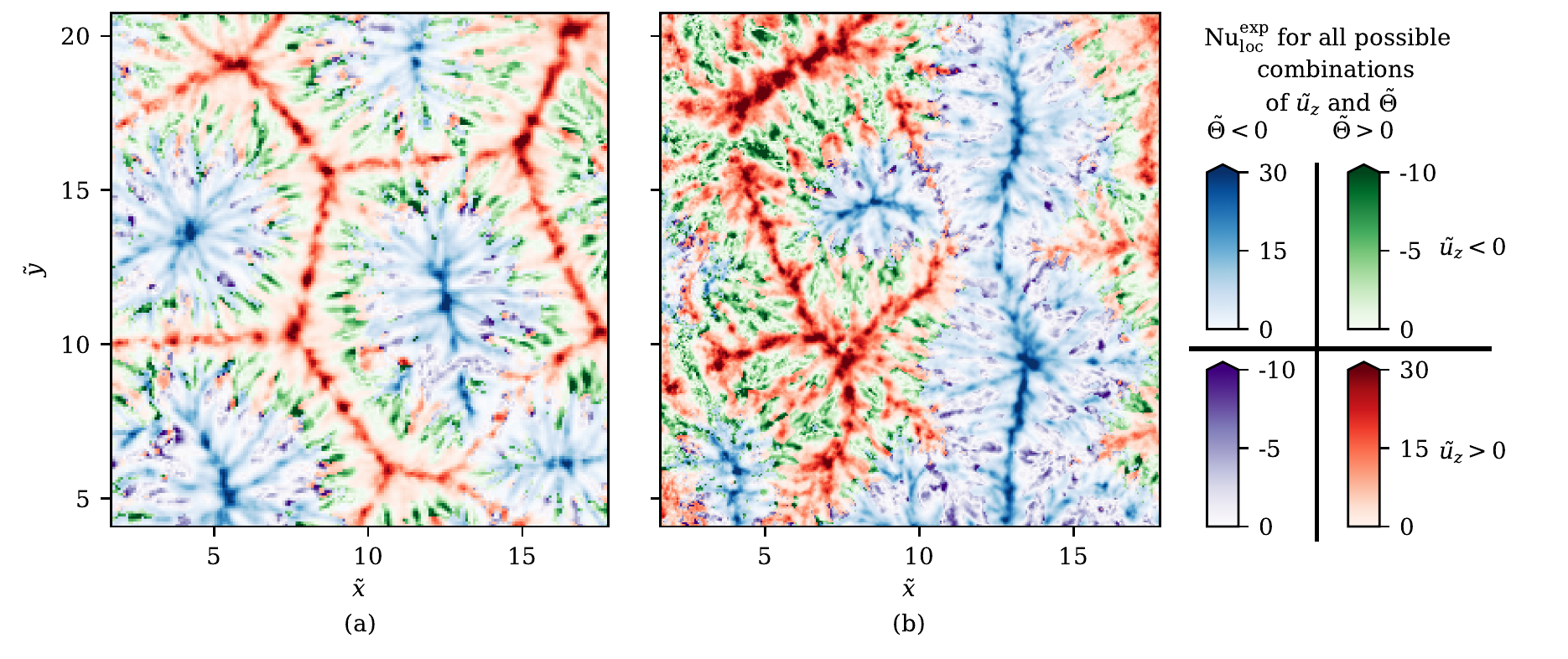}
       \centering
    \caption{Decomposition of $\Nulocexp$ for experimental data at (a) $\Ra = 2 \times 10^5$ and (b) $\Ra = 7 \times 10^5$ depending on the signs of $\tilde{u}_{z}$ and $\tilde{\Theta}$. Both red and blue regions imply $\Nulocexp > 0$, whereas green and purple regions correspond to $\Nulocexp < 0$. The intensity indicates the magnitude -- note here the different scalings for positive and negative $\Nulocexp$. The fields are averaged over 33 $t_{\mathrm{f}}$, which indicates a certain persistence of the inverted heat transfer events. 
    }
    \label{fig:Nusselt_field_decomposition}
\end{figure*}

In order to validate this view, we turn our attention to the entire $\Nulocexp$ fields. Here, we decompose them into the four different possible combinations of $\tilde{u}_z$ and $\tilde{\Theta}$, and visualize exemplary fields averaged over 33$t_{\mathrm{f}}$ for $\Ra = 2 \times 10^5$ and $\Ra = 7 \times 10^5$ in Figure \ref{fig:Nusselt_field_decomposition} (a) and (b), respectively. Note the indication of these different combinations via different colors, whereas their corresponding intensity encodes the magnitude of $\Nulocexp$. As down-welling hot fluid (green) occurs mostly between hot up-welling (red) and cold down-welling (blue) regions, we find our concept of dragged-down hot fluid due to the limited thermal conductivity of the glass plate supported. Since we visualize time-averaged fields, we find these regions of reversed heat transfer to be quite persistent rather than spontaneous. 

Having a closer look at these two panels, we exclusively find ridges of hot up-welling fluid which are mostly connected with each other together with regions of cold down-welling fluid that are surrounded by the former. As outlined by Fonda et al.~\cite{fonda2019deep}, such an arrangement also impacts the heat transfer. The authors categorized the convection pattern based on a deep-learning approach into ridges,  trisectors (connection points of ridges) and wedges (endpoints of ridges). Interestingly, the individual structure classes affect also heat transfer -- for instance, they attributed twice as much heat transfer to trisectors compared to wedges at $\RaD = 10^{5}$. Adopting their terminology to our present configuration, we find several trisectors for hot up-welling fluid but none for cold down-welling fluid -- instead, we only observe wedges for the latter. One might thus expect extended regions of the latter to account for the increased or localized heat transfer of the former. As this is confirmed by our observations, it simultaneously explains why reversed heat transfer due to dragged-down regions appears closer to the hot up-welling regions.
 
Our conceptual sketch consistently explains the far \textit{negative} tails only -- however, returning to Figure \ref{fig:signed_pdf_experiment}, one might wonder why the experimental \textit{positive} tails agree only up to $\Ra = 4 \times 10^{5}$ well but do not match anymore for $\Ra = 7 \times 10^{5}$. Figure \ref{fig:signed_pdf_experiment} (d) shows that these positive far-tail events are dominantly caused by hot (up-welling) fluid. In turn, this implies that cold (down-welling) fluid cannot account for a similar inverse transfer of heat. In fact, this underlines once more the limits that the cooling plate sets on the heat transfer of the entire system and the resulting broader extent of down-welling fluid regions, as elaborated above.

\begin{table}
    \centering
    \begin{tabular}{llc}
        Run & & $\langle \Nulocexp \rangle_{\tilde{A}, \tilde{t}}$ \\ 
    \hline
    DNS     & $\RaD = 1 \times 10^{5}$                              &   $4.04$ \\
    Exp.    & $\Ra \thickspace \thickspace = 2 \times 10^{5}$       &   $4.00$ \\
    Exp.    & $\Ra \thickspace \thickspace = 4 \times 10^{5}$       &   $5.31$ \\
    Exp.    & $\Ra \thickspace \thickspace = 7 \times 10^{5}$       &   $5.58$ \\
    DNS     & $\RaD = 1 \times 10^{6}$                              &   $7.55$
    \end{tabular}
    \caption{Global Nusselt numbers as inferred from their local measurements.
    The time average is taken across the entire corresponding evolution times provided in Figure \ref{fig:structure_size_over_time}. In contrast to the simulations, the latter of which analyze the entire horizontal cross-section, the average across the horizontal plane $\tilde{A}$ is restricted in the experiment to the limited field of view.}
    \label{tab:global_Nu}
\end{table}
Lastly, we turn our focus from the \textit{local} to the \textit{global} heat transfer by computing
$\langle \Nulocexp \rangle_{\tilde{A}, \tilde{t}}$ as summarised in Table \ref{tab:global_Nu} for both the simulations and experiments across the entire $\Ra$-range. Note here that an average across the horizontal plane $\tilde{A}$ is restricted in the experiment to the limited field of view. Nevertheless, the experiment shows a rather good agreement at the lowest Rayleigh number. However, it falls increasingly behind the trend -- that is set by the numerically obtained data for Dirichlet boundary conditions -- as higher $\Ra$ result in successively stronger non-Dirichlet effects at the top plate as explained in Section \ref{Sec:superstructures}. Hence, these effects explain the mismatch of the global Nusselt number when comparing the experiments with numerically obtained results.

\section{Discussion and perspectives}
\label{Sec:Discussion}
In this work, we first investigated the effect of mixed and asymmetric boundary conditions on the size of long-living large-scale flow structures in thermal convection. A comparison of the experimentally observed patterns with numerically obtained patterns for Dirichlet conditions showed an increasing deviation of the former's size from the latter for increasing $\Ra$. We related this trend to a gradual change in the thermal boundary condition at the cooling plate. Despite this difference in the structure size compared to perfect Dirichlet conditions, we confirmed that the patterns in the case of complementary Neumann conditions offered fundamentally different long-term dynamics. In this case, the large-scale flow structures grow gradually until domain size, being in contrast to the fluctuation around a time-independent mean value as observed in the experiments. This confirmed the experimentally obtained structures to be categorized as turbulent superstructures. We underlined our observations based on considerations concerning the linear stability at the onset of convection.

As the thermal boundary conditions vary between the top and bottom plate -- and change in the case of the former even with $\Ra$ --, the primary instability is similarly affected. However, the final flow establishes a balance between both conditions. To investigate the effect of potential slight variations of the patterns over the layer height, further experiments are necessary. These would need to incorporate a simultaneous and fully volumetric temperature and velocity measurement across the entire height since the temperature gradients must not be neglected within the thermal boundary layers. 
Motivated by the infinite growth of critical patterns at the onset of convection in the Neumann case, it would further be of interest at which critical ratio of thermal diffusivities such behavior sets in.

The second part of our work addresses the influence of the experimentally asymmetric boundary conditions on local and global heat transfer. From a systematic derivation based on high-resolution numerical data, we found that the coarse horizontal resolution of the experimental measurement affects the statistics of the local Nusselt number most strongly. 
Contrasting the probability density functions of the local Nusselt number $\Nulocexp$ from experimental and numerical data signalized higher probabilities of inverted heat transfer for the experiment. A subsequent decomposition of $\Nulocexp$ depending on the sign of the temperature deviation $\Theta$ unveiled an asymmetry due to the limited thermal conductivity of the cooling plate. In other words, the cooling plate falls short of generating large magnitude $\Nuloc$ events or pronounced cold plumes. We confirm this interpretation by detailed horizontal cross-sections of the local Nusselt number, which exhibits extended regions of hot down-welling fluid located between regions of hot up-welling and cold down-welling fluid.

Our study highlights the importance of systematic investigations of convection with mixed or asymmetric boundary conditions beyond the linear stability. Hence, engineers and scientists are asked to develop advanced measurement techniques for combined volumetric temperature and velocity measurements in laboratory experiments -- this would eventually allow for long-term investigations of the slow dynamics of turbulent superstructures as well as other long-living large-scale flow structures and the influence of asymmetric thermal boundary conditions that go beyond the present possibilities of the present studies.

\section*{Acknowledgement}
The authors gratefully acknowledge Sebastian Moller and Ambrish Pandey for providing the experimental measurement and numerical data, respectively. This work was supported by the DFG Priority Programme SPP 1881 on \enquote{Turbulent Superstructures}. The work of T.K. was partly funded by the Carl Zeiss Foundation within project no. P2018-02-001 \enquote{Deep Turb – Deep Learning in and of Turbulence}.



\bibliographystyle{elsarticle-num} 
\bibliography{main}





\end{document}